\documentclass[runningheads]{llncs}
\usepackage{graphicx}
\usepackage{multirow}
\usepackage{amsmath}
\usepackage{amsfonts}
\usepackage{subcaption}
\usepackage{blindtext}

\usepackage[dvipsnames]{xcolor}

\begin{document}
\title{Analyzing the Effects of Handling Data Imbalance on Learned Features from Medical Images by Looking Into the Models} 
%
\titlerunning{Effects of Handling Data Imbalance on Learned Features}

\author{Ashkan Khakzar\textsuperscript{1}\thanks{denotes equal contribution},
Yawei Li\textsuperscript{1*},
\\
Yang Zhang\textsuperscript{1},
Mirac Sanisoglu\textsuperscript{1},
Seong Tae Kim\textsuperscript{2},
Mina Rezaei\textsuperscript{3},
\\
Bernd Bischl\textsuperscript{3},
Nassir Navab\textsuperscript{1}}
%
\authorrunning{ }
%
\institute{\textsuperscript{1}Technical University of Munich (TUM) \\
\textsuperscript{2}Kyung Hee University\\
\textsuperscript{3}University of Munich (LMU)}

\maketitle              

\begin{abstract}

One challenging property lurking in medical datasets is the imbalanced data distribution, where the frequency of the samples between the different classes is not balanced. Training a model on an imbalanced dataset can introduce unique challenges to the learning problem where a model is biased towards the highly frequent class. Many methods are proposed to tackle the distributional differences and the imbalanced problem. However, the impact of these approaches on the learned features is not well studied. In this paper, we look deeper into the internal units of neural networks to observe how handling data imbalance affects the learned features. We study several popular cost-sensitive approaches for handling data imbalance and analyze the feature maps of the convolutional neural networks from multiple perspectives: analyzing the alignment of salient features with pathologies and analyzing the pathology-related concepts encoded by the networks. Our study reveals differences and insights regarding the trained models that are not reflected by quantitative metrics such as AUROC and AP and show up only by looking at the models through a lens.
\footnote[1]{\url{https://github.com/CAMP-eXplain-AI/imba-explain}}

\end{abstract}




\vspace{-4pt}
\section{Introduction} ~\label{sec:intro}
Medical imaging datasets often appear in imbalanced distribution where the frequency of the samples between the different classes of the training dataset is not similar or balanced. The low amount of training samples for infrequent classes or tailed distribution makes it challenging to learn optimal classification boundaries in the representation and can lead to a biased model. 
Existing methods to tackle class imbalance problem either modify data distribution or learn appropriate costs to re-weight class errors. At the data level, the objective is to balance the data distribution through re-sampling techniques which often are prone to over-fitting~\cite{khan2019striking,buda2018systematic,he2009learning}. On the other hand, the cost-sensitive approaches modify the learning algorithm to alleviate the bias towards frequent classes or head of distribution~\cite{cui2019class,rezaei2018generative}. The efficacy of these approaches is demonstrated by conventional metrics such as precision and recall and their derivatives. However, the effect of these approaches on the learned features is not well studied. The study of learned features not only improves our understanding of what happens within the models, is specifically insightful when the conventional evaluation metrics do not reflect the effect of applying such cost-sensitive approaches.

This study explores what happens to the learned features when it is trained by cost-sensitive approaches to handle the data imbalance. To understand the effect on learned features, we analyze the internal units of neural networks. Specifically, we analyze the feature maps (outputs of convolutional layers) in deep layers by class activation maps and network dissection~\cite{bau2017network}.
First, we optimize models (ResNet \cite{he2016deep} and DenseNet \cite{huang2017densely}) with Binary Cross-Entropy (BCE)~\cite{murphy2012machine}. Several recent and popular cost-sensitive losses such as Weighted BCE (WBCE)~\cite{paszke15pytorch}, Focal loss~\cite{lin2017focal}, and Class-Balanced Focal loss (CB-Focal)~\cite{cui2019class} on NIH Chest X-ray \cite{wang2017chestx} dataset and report classical metrics: area Under ROC curve (AUROC), average precision (AP) and predicted probabilities. Then, we visually analyze the impact on salient learned features using class activation maps \cite{zhou2016learning,selvaraju2017grad} and provide quantitative evaluations to validate the visual observation. We then proceed to analyze the learned features using network dissection~\cite{bau2017network,khakzar2021towards} which quantitatively identifies the concepts encoded (learned) by the model.

\noindent\textbf{Statement of Contribution}
By placing models trained with different learning strategies under the lens, we observe that while metrics such as AUROC and AP report equivalent performance, the models trained with cost-sensitive losses encode more pathology-related concepts. Moreover, we observe an increased alignment between salient learned features and pathology-related features.

\vspace{-4pt}
\section{Related Works} ~\label{sec:relatedworks}
\textbf{Handling Data Imbalance:} Much of the recent works on imbalanced learning focused on alleviating this problem using novel objective function. Lin \textit{et. al.}~\cite{lin2017focal} introduce a Focal loss for dense object detection where class-specific weights were automatically learned. Cui \textit{et. al.}~\cite{cui2019class} re-weight the loss by the inverse effective number of examples to learn balanced representations. Similarly, ~\cite{khan2019striking} modified the weights according the uncertainty of predictions. 
Others address this problem by multi-task learning~\cite{kendall2018multi,rezaei2018generative} that used selective instances for training on imbalanced sets for each task.
\\
\textbf{Interpreting Neural Networks:} Two principal neural network interpretation approaches are feature attribution~\cite{simonyan2013deep,zhang2021fine,Khakzar_2021_CVPR,Sundararajan2017,Lundberg2017,khakzar2021explaining,khakzar2019improving} (i.e. saliency methods~\cite{cong2018review}) and analyzing internal units (e.g. feature visualization~\cite{olah2017feature} and dissection~\cite{bau2020understanding,khakzar2021towards}). Here, we look into the models from both perspectives. We are interested in both the contribution of input features to the output (i.e., feature attribution) and concepts encoded by the network (via analyzing units). %
The are many attribution methods, %
however the identified important features are different for different methods \cite{krishna2022disagreement,zhang2021fine,khakzar2022explanations,khakzar2020rethinking}. This is a caveat for researchers using the attribution toolkit.
Within the various feature attribution methods, CAM~\cite{selvaraju2017grad} being a classic and intuitive method, it satisfies benchmarks in terms of faithfulness \cite{hooker2019benchmark,zhang2021fine,khakzar2022explanations}. Moreover, regardless of its advantages for attribution, the method provides a summary of activation maps in a certain layer by performing a weighted sum of the attributions. 
We use network dissection \cite{bau2017network,khakzar2021towards} to identify concepts associated with individual convolutional feature maps.
%






\vspace{-2pt}
\section{Method} ~\label{sec:method}
\vspace{-2pt}
The question we explore in this paper is what happens to the learned features within the model when we apply cost-sensitive approaches to handle the class imbalance during training. Specifically, we study the intra-class data imbalance, i.e. the imbalance between the number of positive and negative samples for each class. We first introduce the cost-sensitive methods (see Section~\ref{sec:method:imb}) that consider the data imbalance between positive and negative samples. 
Then,  we explain our approaches for analyzing the learned features (see Section~\ref{sec:ourmethod}).
\vspace{-4pt}
\subsection{Handling Data Imbalance} \label{sec:method:imb}
Given $\mathcal{D} = \{x^{(k)}, y^{(k)} \}_{k=1}^K$, our goal is to learn the optimal boundary ${\theta}^*$ obtained by empirical loss minimization $\mathcal{L}_\mathcal{D} (\theta)$ on the training set $\mathcal{D}$ as: $
{\theta}^* =  \arg \min_{\theta}  \mathcal{L}_{\mathcal{D}} (\theta)$. The class imbalanced problem exists when the frequency of the samples among different categories are extremely mismatched. Therefore $\theta$ learned on $\mathcal{D}$ using
conventional empirical loss can be biased towards the low frequent classes and significantly different from the ideal boundary $\theta^*$. In other words, because of the imbalanced proportion between classes, the optimal boundary is more likely to afford a higher empirical error than an alternative hypothesis based on an empirical loss. This is due to the nature of imbalanced class distribution that forces the classifier to shift $\theta$ closer to low-frequent classes because it reduces empirical error. A principal strategy to handle data imbalance is through the cost-sensitive loss functions~\cite{lin2017focal,cui2019class,kendall2018multi,ridnik2021asymmetric}. 

Here, we study data imbalance problem in multi-label and binary classification. Assume $N$ samples $\mathcal{X} = \{x^{(i)}\}_{i=1}^N$ from $M$ classes. we denote the corresponding label set as $\mathcal{Y} = \{y^{(i)}\}_{i=1}^N$ with $y^{(i)} \in \{1,2,...,M\}$. The probability $p_m^{(i)}$ of class $m$ for sample $x^{(i)}$ given by a neural network is defined with multiple outputs as cross entropy error between predicted value and true label using Binary Cross Entropy (BCE) loss. The cost-sensitive losses applied to class can be summarized into a unified framework:
\begin{equation} \label{eq_1}
   \mathcal{L_{CE}} = - \sum_{i=1}^{N} [w_{+}^{(i)} y^{(i)}\log p^{(i)} + w_{-}^{(i)} (1 - y^{(i)}) \log (1 - p^{(i)})],
\end{equation}
where $w_{+}^{(i)}$ and $w_{-}^{(i)}$ are the positive and negative class weights for sample $x^{(i)}$, respectively. We omit the class index $m$ for simplicity. The four losses considered in this work can be derived as follows:
\begin{equation}
    \begin{aligned}
        \text{BCE~\cite{murphy2012machine}:} \quad & w_{+}^{(i)} = w_{-}^{(i)} = 1, \\
        \text{WBCE~\cite{paszke15pytorch}:} \quad & w_{+}^{(i)} = \frac{N_{-}}{N_{+} + N_{-}}, \ \text{and} \ w_{-}^{(i)} = \frac{N_{+}}{N_{+} + N_{-}}, \\
        \text{Focal~\cite{lin2017focal}:} \quad & w_{+}^{(i)} = \alpha (1 - p^{(i)})^{\gamma}, \ \text{and} \ w_{-}^{(i)} = (1 - \alpha) {p^{(i)}}^\gamma, \\
        \text{CB-Focal~\cite{cui2019class}:} \quad & w_{+}^{(i)} = \frac{1 - \beta}{1 - \beta^{N_{+}}} \cdot (1 - p^{(i)})^\gamma, \ \text{and} \ w_{-}^{(i)} = \frac{1 - \beta}{1 - \beta^{N_{-}}} \cdot {p^{(i)}}^\gamma, \\
    \end{aligned}
\end{equation}
where $N_{+}, N_{-}$ are the number of positive and negative samples of class $m$, and $\alpha, \gamma$ and $\beta$ are hyper-parameters. 
As we are discussing intra-class imbalance, each loss is considered individually and we apply the cost-sensitive methods on each loss to strike a balance between positive and negative samples for each class.

\vspace{-4pt}
\subsection{Looking into the Models} \label{sec:ourmethod}
The objective is to analyze the features learned by the model in different training scenarios.
In this work, we analyze the internal units of the model, specifically the activation pattern of convolutional feature maps in deeper layers. The deep convolutional layers are chosen due to the following two reasons: 1) Final layers in principle encode higher-level concepts and we are interested to know if these concepts align with pathological features. 2) We focus on convolutional layers instead of deeper fully connected layers, since convolutional layers keep the spatial correspondence with input features, and therefore, we can leverage image annotations from experts to study if the activations align with pathology related features.
%
We analyze the internal features maps from two different perspectives: 

\textbf{Class Activation Maps \cite{zhou2016learning,selvaraju2017grad}} can be deemed as a tool that summarizes the contributions of feature maps of the final convolutional layer in one tensor (for the networks under investigation, GradCAM \cite{selvaraju2017grad}, and CAM \cite{zhou2016learning} are equivalent). The method combines the feature maps using their associated weight (or gradient of output with respect to feature map in GradCAM) of their connections to the output, thus summarizing how feature maps contribute to the final output. The method, in essence, shows which areas are contributing to the output for each sample input. Thus using annotations from experts, we can check whether the salient regions are aligned with the pathologies \cite{wang2017chestx,khakzar2019learning}.

\textbf{Analyzing Encoded Concepts \cite{bau2017network,khakzar2021towards}:}  We use an approach inspired by network dissection~\cite{bau2017network} which is a tool for individually analyzing the convolutional feature maps of the neural network. In essence, it identifies the concepts associated with each feature map, thus allowing for understanding what concepts are encoded by the network during training. In this work, 
we use a conceptually similar approach. For a chosen convolutional layer, let $A^{c}$ denote the activation (feature) map of channel $c$. Similar to~\cite{bau2017network}, we first apply a threshold $\tau_{c}$ with $\mathbb{P}(A^C \geq \tau_c) = q$ to activation values of each channel $c$ to filter the dataset-wise significant activation values, where $q$ is a pre-defined hyper-parameter. For each image in the annotation dataset we compute the corresponding $A^{c}$ and apply the threshold $\tau_{c}$. Subsequently, we identify the connected components in the threshold activations. If the connected component overlaps the bounding box region, we consider the connected component as a pathology related concept. Using this approach we can report two values that reflect the concepts encoded in the network quantitatively, which are explained in Sec. \ref{sec:experiments}.

\vspace{-4pt}
\section{Experiments} \label{sec:experiments}
\vspace{-2pt}
\subsubsection{Experimental Setup}
In this paper, we examine and perform two different datasets and learning tasks: multi-label classification and binary classification on the NIH Chest X-ray dataset~\cite{wang2017chestx}. For binary classification. We define samples with the ``No Finding'' label as ``Healthy'' samples, and samples with other labels as ``Unhealthy'' samples. 
Random crop, horizontal flip, and color jitter are used as the augmentations. In the end, the augmented images are resized to $224 \times 224$ before being fed into an image classifier.

We use two ImageNet pre-trained classifiers: ResNet50~\cite{he2016deep} and DenseNet~\cite{huang2017densely}. The last feature maps of ResNet50 have the size $7 \times 7$. In order to have a fine-grained analysis on smaller features such as Nodule (for which $7 \times 7$ is a coarse resolution) we use a truncated version of DenseNet. We discard the last two dense blocks and the associated transition blocks, so that the last feature maps have size $28 \times 28$. We refer to this truncated variant of DenseNet as \textit{T-DenseNet}.
We adopt the same training configurations for all models. Specifically, we train the classifiers on 4 GPUs (DGX-A100) for 50 epochs using the Adam~\cite{diederik2015adam} optimizer with weight decay $10^{-6}$ and initial learning rate $0.0004$. The learning rate is decayed following the cosine-annealing policy. In addition, we set the batch size to 512.
For the Focal loss, the $\gamma$ and $\alpha$ set to 2.0 and 0.25, respectively; for the class-balanced focal loss, we set $\gamma$ to 2.0 and $\beta$ to 0.9999.

\begin{figure}[!t]
  \centering
  \subfloat[]{\includegraphics[width=0.82\textwidth]{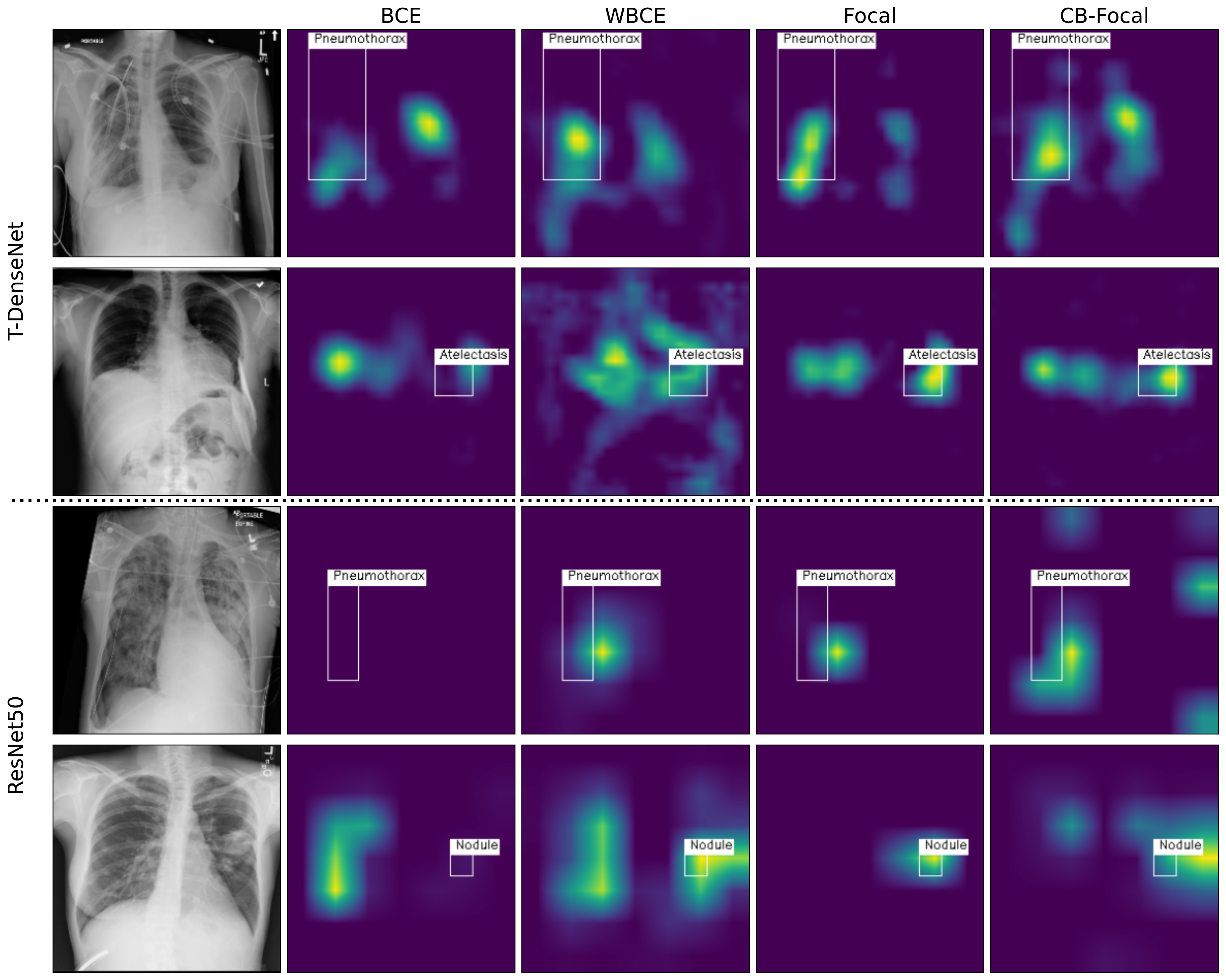}}
  \hfill
  \subfloat[]{\includegraphics[width=0.82\textwidth]{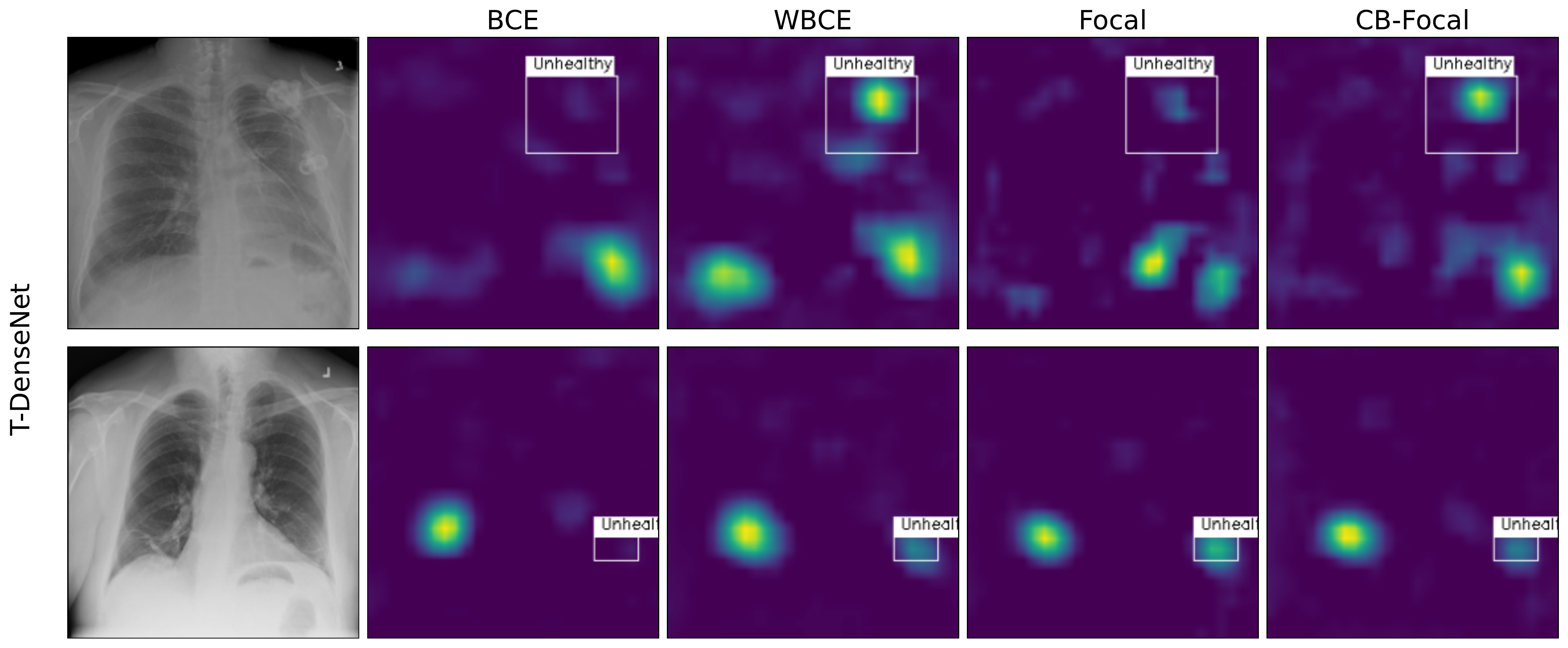}\label{fig_2b}}
  \vspace{-6pt}
  \caption{\textbf{Feature Alignment with Pathologies.} (a) multi-label classification task (b) binary classification task --
  Each row shows CAMs of different models on a single sample (the heatmap values are normalized to $[0, 1]$). Each column is associated with a training strategy (BCE, WBCE, Focal, CB-Focal). We observe that increased activation values due to training with cost-sensitive approaches appear primarily in areas where the pathologies exist. To quantitatively show this effect on the entire dataset, we report IoBB values in Tab.~\ref{tab:iobb_ior}.}
\label{fig:gradcams}
\end{figure}

\vspace{-4pt}
\subsubsection{Feature Alignment Analysis}
The objective is to have a metric for measuring the alignment between the contributing features (from CAMs) for a single prediction and the pathologies. This can be achieved by classical object detection metrics IoBB(Intersection over Bounding Box) and IoR(Intersection of detected Region). IoBB and IoR can be considered as the visual counterparts of recall and precision. Computing these metrics requires thresholding the CAMs. In order to reduce the sensitivity of the results to the chosen threshold, we implement soft IoBB and IoR computation. This is performed by normalizing the CAMs to $[0, 1]$ and applying a summation of values within the regions under consideration (intersection, bounding box, detected region).
\vspace{-3pt}
\subsubsection{Analysis of Concepts}
We count the number of concepts within bounding boxes in two ways: \textbf{Disjoint:} Number of concepts detected by a feature map (i.e. connected components overlapping the bounding box) for all feature maps in the chosen layer and all images in the annotation subset and normalize the value by the number of images. This approach considers repeated concepts in a bounding box, as some bounding boxes cover huge regions where multiple concepts occur. \textbf{Unique:} For each feature map, if there is at least one concept detected, we consider the feature map as a unique concept detector (similar to~\cite{bau2017network}). We choose the hyper-parameter $q=0.01$ for the T-DenseNet and $q=0.04$ for the ResNet50. Different $q$ only changes the absolute number of detected concepts and does not affect comparative analysis. We count the number of unique concept detectors for all images and normalize by the number of images. The provided metrics reflect how individual feature maps are aligned with pathologies.


\vspace{-6pt}
\section{Results and Discussion} \label{sec:results}
\vspace{-2pt}
\subsubsection{Performance Analysis}
\begin{figure}[!t]
    \centering
    \includegraphics[width=\textwidth]{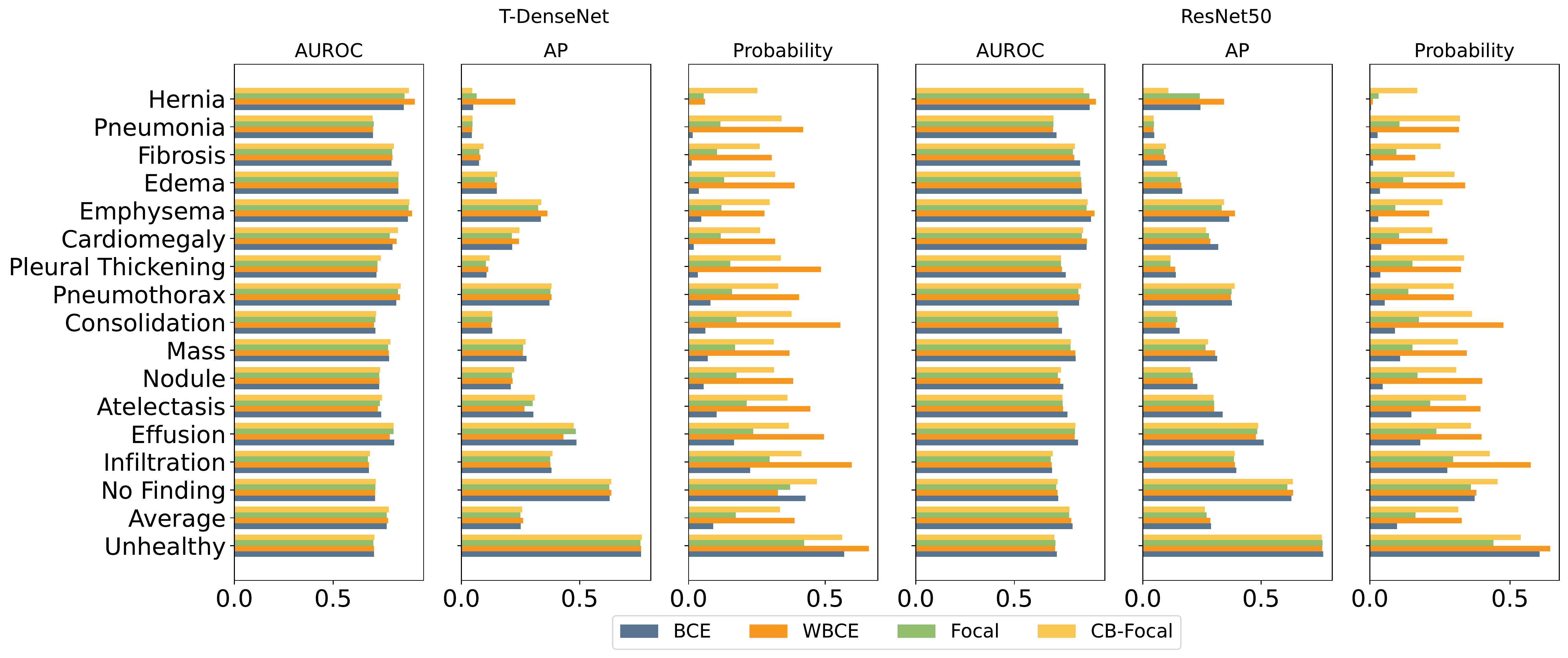}
    \caption{\textbf{Performance Analysis.} The sub-figures $1 \sim 3$ (from left to right) and $4 \sim 6$ show the results of the T-DenseNet and ResNet50, respectively. In addition to AUROC and AP, sample-wise averaged predicted probabilities are shown. In each sub-figure, ``Average'' indicates the class-wise average of the corresponding metric in multi-label classification;  ``Unhealthy'' refers to the positive class in binary classification task. We observe that AUCROC and AP for different learning strategies are similar. Thus these metrics do not tell the entire story.}
    \label{fig:class_performance}
    \vspace{-4pt}
\end{figure}

We evaluate the classification performance using AUROC and AP
(Fig.~\ref{fig:class_performance}). The performance differences between BCE and weighted losses (WBCE, Focal, and CB-Focal) are marginal. Therefore the effect of handling data imbalance via losses is not visible via these metrics. We also report the predicted probabilities of different losses. The predicted probabilities of the weighted losses are substantially higher than that of the BCE. This is expected as more weight is given to positive samples. Moreover, as more weight is given to the positive weights, the recall is higher. However, it is not clear if the increased recall is due to model's inclination towards identifying samples as positive or due to the model having learned and leveraging more predictive features.

\vspace{-4pt}
\subsubsection{Feature Alignment Analysis}
In this section, we analyze the models' internal feature maps to see if the increased recall and probability values are related to predictive features relevant to the pathologies. This is realized by comparing CAMs and the salient features for individual predictions with annotations from radiologists. In Fig. \ref{fig:gradcams} we observe that in models trained with cost-sensitive losses, the additionally activated regions are aligned with the pathologies (bounding boxes). This shows that the increased recall is not due to the models' propensity towards identifying samples as positive but due to having learned meaningful predictive features. Note that metrics such as AUROC and AP are roughly equivalent for all models (Fig. \ref{fig:class_performance}). In order to quantify the observed behavior in Fig. \ref{fig:gradcams} we report IoBB and IoR results in Tab. \ref{tab:iobb_ior}. The increased IoBB for models trained with cost-sensitive losses shows that the features cover more areas of bounding boxes (pathologies). We also observe a decrease in IoR, which is also evident in Fig. \ref{fig:gradcams} as falsely positive active regions have also increased. 

\begin{table}[t]
    \centering
    \begin{tabular}{c | c c| c c | c c| c c}
    \hline
    Task & \multicolumn{4}{c|}{Multi-label} &  \multicolumn{4}{c}{Binary}\\
    \hline
    Classifier & \multicolumn{2}{c|}{T-DenseNet} & \multicolumn{2}{c|}{ResNet50} & \multicolumn{2}{c|}{T-DenseNet} & \multicolumn{2}{c}{ResNet50} \\
    \hline
    Metric & IoBB & IoR & IoBB & IoR & IoBB & IoR & IoBB & IoR \\
    \hline
    BCE & 0.2105 & 0.2951 & 0.2700 & \textbf{0.2401} & 0.2661 & \textbf{0.2123} & 0.3159 & \textbf{0.2084} \\
    WBCE & \textbf{0.2836} & 0.2252 & \textbf{0.3447} & 0.2060 & \textbf{0.2765} & 0.1999 & \textbf{0.3505} & 0.2021 \\
    Focal & 0.1855 & \textbf{0.3223} & 0.2915 & 0.1845 & 0.2129 & 0.2064 & 0.2516 & 0.2042 \\
    CB-Focal & 0.2753 & 0.3141 & 0.2839 & 0.2375 & 0.2458 & 0.1956 & 0.3409 & 0.1955 \\
    \hline
    \end{tabular}
    \caption{\textbf{Feature Alignment Analysis.} The average IoBB and IoR results for test sets images that have bounding box annotations. These metrics are provided to quantitatively validate the observation in Fig. \ref{fig:gradcams} for the entire test set.}
    \label{tab:iobb_ior}
    \vspace{-4pt}
\end{table}

\vspace{-6pt}
\subsubsection{Analysis of Concepts}

This section analyzes the effect of handling data imbalance on the concepts encoded by the model. This is performed by counting the number of detected pathology-related concepts. In Tab. \ref{tab:concepts} we observe that the models trained with cost-sensitive losses consistently have more detectors. Thus it can be concluded that the models encode more pathology-related concepts when the data imbalance is considered in the loss.
\begin{table}[t]
    \centering
    \begin{tabular}{c | c c| c c | c c| c c}
    \hline
    Task & \multicolumn{4}{c|}{Multi-label} &  \multicolumn{4}{c}{Binary}\\
    \hline
    Classifier & \multicolumn{2}{c|}{T-DenseNet} & \multicolumn{2}{c|}{ResNet50} & \multicolumn{2}{c|}{T-DenseNet} & \multicolumn{2}{c}{ResNet50} \\
    \hline
    Metric & Disjoint & Unique & Disjoint & Unique & Disjoint & Unique & Disjoint & Unique \\
    \hline
    BCE & 217 & 141 & 562 & 328 & 280 & 165 & 363 & 278 \\
    WBCE & 259 & 157 & \textbf{641} & 441 & 287 & 169 & 542 & 386\\
    Focal & 226 & 143 & 585 & \textbf{570} & 291 & 170 & \textbf{586} & \textbf{414} \\
    CB-Focal & \textbf{263} & \textbf{163} & 546 & 390 & \textbf{292} & \textbf{171} & 567 & 400 \\
    \hline
    \end{tabular}
    \caption{\textbf{Analysis of Concepts.} The normalized number of detected concepts (Unique and Disjoint) in the chosen convolutional layers of DenseNet and ResNet models for the two tasks. Handling data imbalance consistently increases the number of concepts encoded by the model.}
    \label{tab:concepts}
\end{table}


%


\vspace{-5pt}
\section{Conclusion} \label{sec:conclusion}
\vspace{-4pt}
In this work, we study the effect of handling data imbalance using cost-sensitive losses during training on the learned feature maps. The feature maps are analyzed from two perspectives: class activation maps and the concept encoded by each feature map. We observe that although classical metrics such as AUROC and AP report equivalent performance for the studied models, the learned features are different in these models. When data imbalance is handled, the models encode more pathology-related concepts. Moreover, we observe that overall increased recall and predicted probability of these models are accompanied by an increased alignment between learned features and pathology-related features.

\paragraph{Acknowledgment}
This work has been funded in part by the German Federal Ministry of Education and Research (BMBF) under Grant No.01IS18036A and 01IS18036B, Munich Center for Machine Learning (MCML). A. K., N. N. were supported with funding from the Bundesministerium fur Bildung und Forschung (BMBF). M. R., and B. B. were supported by the German Federal Ministry of Education and Research (BMBF) under Grant No. 01IS18036A. Y. L. is supported by the German Federal Ministry of Health (2520DAT920). S.T. K. is supported by National Research Foundation of Korea(NRF) grant funded by the Korea Government(MSIT) (No. 2021R1G1A1094990).

%
%
 \bibliographystyle{splncs04}
 \bibliography{ref}
 
\end{document}